\documentclass[12pt,preprint]{aastex}
\usepackage{pifont}
\usepackage{amssymb}

\shorttitle{Jitter Self-Compton Processes of GRB}

\shortauthors{Mao \& Wang}

\begin{document}


\title{Jitter Self-Compton Process: GeV Emission of GRB 100728A}


\author{Jirong Mao}
\affil{Korea Astronomy and Space Science Institute, 776,
Daedeokdae-ro, Yuseong-gu, Daejeon 305-348, Republic of Korea}
\affil{Yunnan Observatory, Chinese Academy of Sciences, Kunming,
Yunnan Province 650011, China} \affil{Key Laboratory for the
Structure and Evolution of Celestial Objects, Chinese Academy of
Sciences, Kunming, China}

\email{jirongmao@kasi.re.kr}

\author{Jiancheng Wang}
\affil{Yunnan Observatory, Chinese Academy of Sciences, Kunming,
Yunnan Province, 650011, China} \affil{Key Laboratory for the
Structure and Evolution of Celestial Objects, Chinese Academy of
Sciences, Kunming, China}

\begin{abstract}
Jitter radiation, the emission of relativistic electrons in a random
and small-scale magnetic field, has been applied to explain the
gamma-ray burst (GRB) prompt emission. The seed photons produced
from jitter radiation can be scattered by thermal/nonthermal
electrons to the high-energy bands. This mechanism is called jitter
self-Compton (JSC) radiation. GRB 100728A, which was simultaneously
observed by the {\it Swift} and {\it Fermi}, is a great example to
constrain the physical processes of jitter and JSC. In our work, we
utilize jitter/JSC radiation to reproduce the multiwavelength
spectrum of GRB 100728A. In particular, due to JSC radiation, the
powerful emission above the GeV band is the result of those jitter
photons in X-ray band scattered by the relativistic electrons with a
mixed thermal-nonthermal energy distribution. We also combine the
geometric effect of microemitters to the radiation mechanism, such
that the ``jet-in-jet" scenario is considered. The observed GRB
duration is the result of summing up all of the contributions from
those microemitters in the bulk jet.
\end{abstract}




\section{Introduction}
Gamma-ray bursts (GRBs) are objects emitting high-energy photons.
One detection of a GeV photon from GRB 940217 was reported 17 years
ago \citep{hurley94}. Compared with the previous research, recently,
with Large Area Telescope (LAT) on board the {\it Fermi} satellite,
we have more observational cases for the study of GRB high-energy
emission above 100 MeV. It is more important that the published
multi-wavelength data of GRB 090510 and GRB 100728A have been
provided by the simultaneous observations of the {\it Swift} and
{\it Fermi} \citep{depasquale10,abdo11}. Some radiation models can
be constrained by these simultaneous data.

It is hard to apply the simple synchrotron model for the explanation
of high-energy emission from GRB 941017 \citep{gonzalez03}. In
general, the photons produced by synchrotron radiation can be
scattered by the relativistic electrons. Therefore, an inverse
Compton process or synchrotron self-Compton (SSC) process is
proposed naturally to explain the GRB emission above the GeV band
\citep{meszaros93,dermer00,wang01,zhang01,granot03,fan08,zou09,corsi10}.
In particular, the possibility was suggested that the photons from
X-ray flares can be scattered to the GeV band by those relativistic
electrons \citep{wang06,galli08}.

From a theoretical point of view, \citet{meszaros94} first mentioned
the physics of turbulent field growth for the study of GRB
radiation. \citet{narayan09} proposed one model in which the GRB
radiating fluid is relativistically turbulent. This turbulent
process, plus the inverse Compton mechanism, was applied to the
study of radiation in GRB 080319B \citep{kumar09}. In the turbulent
fluid, the random and small emitters can produce short-time
variabilities, indicating many pulses shown in the GRB prompt light
curve \citep{lyutikov06,lazar09}. It is worth noting the key point
of this ``jet-in-jet" model: these microemitters within the bulk jet
of GRB explosion also have a jet structure.

The turbulent scenario mentioned above is consistent with the
principle of jitter mechanism. Jitter radiation, which is the
emission of relativistic electrons in a random and small-scale
magnetic field, has been applied to GRB research \citep{medvedev00,
medvedev06}. The random and small-scale magnetic field can be
generated by Weibel instability \citep{medvedev99}. Alternatively,
we propose that the turbulent cascade process can also produce the
random and small-scale magnetic field \citep{mao07,mao11}. As the
magnetic field may have a sub-Larmor length scale, the jitter
radiation in this sub-Larmor scale magnetic field was fully studied
by \citet{med09} and \citet{med11}. The small-scale turbulent dynamo
with large Reynolds numbers at a saturated state in a fluid flow was
simulated by \citet{sch04}. The simulation identified a power-law
turbulent energy spectrum. In our model, the complicated jitter
radiation is simplified as a one-dimensional case to study GRB
prompt emission. From this specified jitter radiation, the spectral
index of a single electron is directly related to the turbulent
energy spectrum.

In general, the electron energy distribution is assumed to be a
power law. However, in the turbulent framework, stochastic
acceleration may be effective.
\citet{schlickeiser89a,schlickeiser89b} found a Maxwellian energy
distribution of electrons. A similar quasithermal stationary
solution of stochastic acceleration was given by \citet{ka06}. In a
turbulent magnetic field, the stochastic acceleration of
ultrarelativistic electrons was also discussed by \citet{stawarz08}.
With the Maxwellian electron energy distribution, the radiative
spectrum and light curve of GRB afterglow were calculated by
\citet{giannios09}.

As suggested by \citet{kirk10}, if the jitter photons in the keV
band are scattered by the relativistic electrons, the final output
emission will be in the GeV band. In this work, we attempt to
calculate the inverse Compton scattering of jitter radiation.
Similar to the SSC mechanism, this process can be called as ``jitter
self-Compton" (JSC) mechanism. At present, we have only two
published data sets of GRBs (GRB 090510 and GRB 100728A), which were
obtained by the simultaneous observations from {\it Swift} and {\it
Fermi}. In particular, the extremely powerful X-ray flares and GeV
emission of GRB 100728A were observed by {\it Swift}/X-ray telescope
(XRT) and {\it Fermi}/LAT, respectively. Thus, the case of GRB
100728A provides us an excellent chance to study the powerful GeV
emission and the link between keV emission and GeV emission in the
GRB multiwavelength spectrum. In this work, the multiwavelength
spectral result of GRB 100728A is especially precious to constrain
our theoretical model of jitter/JSC radiation. Furthermore, from the
clues of \citet{lyutikov06} and \citet{lazar09}, we expect that the
observed gross emission from a bulk jet launched by GRB explosion
might be related to the emissions from the small-scale emitters with
the minijet structure. Therefore, in this paper, we stress the
following issues. (1) To prove the jitter/JSC mechanism, which may
work for GRB prompt/GeV-band emission, we use the multiwavelength
spectrum of GRB 100728A \citep{abdo11} as an example. (2) In our
former research \citep{mao11}, jitter has been identified as the
possible radiation mechanism for the GRB prompt emission, and
turbulence is the dominant dynamic process. Here, some further,
detailed calculations are required to fit the observational data of
GRB 100728A. (3) The JSC mechanism can be examined because the
multiwavelength spectrum of GRB 100728A has been given. (4) The
final JSC result is dependent on the electron energy distribution.
(5) Since the turbulent dynamics has been applied in \citet{mao11},
as a consequent step, the link between the microemitters with the
minijet structure and the emission of the GRB bulk jet should be
considered. (6) As calculated by \citet{mao11}, the cooling
timescale of relativistic electrons has a typical value of about
$10^{-8}$ s. The observed duration of GRB prompt emission is much
longer than the cooling timescale, so further explanations are
required.

In Section 2.1, we first briefly describe our specific case of
jitter radiation and then fully present the JSC process. In Section
2.2, we illustrate the ``jet-in-jet" scenario and link the minijets
to the bulk jet structure. In Section 2.3, combined with the
``jet-in-jet" effect, our jitter/JSC model can reproduce the
multiwavelength spectral properties of GRB 100728A. In particular,
we focus on the GeV emission detected by {\it{Fermi}}/LAT and the
JSC process. The observed GRB duration can also be estimated.
Conclusions and discussion are given in Section 3.

\section{JSC Process and Application of GRB 100728A}
There are two important issues concerning the following jitter/JSC
calculations: (1) Because stochastic acceleration is one of the key
points in this work, the Maxwellian energy distribution of
relativistic electrons should be applied in the calculation. (2) The
gross temporal profile of GRB prompt emission can be the result of
superimposing from a large amount of short-timescale pulses.

\subsection{JSC Process}
Jitter radiation is the emission of relativistic electrons in a
random and small-scale magnetic field. In a one-dimensional case,
the simplified formulas of jitter radiation have been derived
\citep{mao07,mao11}. We also propose that a random and small-scale
magnetic field can be produced by turbulence. The radiative
intensity of a single electron has a power-law shape, and the
spectral index is related to the energy spectrum of turbulent flow.
Here, we write the radiative intensity in the unit of
$\rm{erg~s^{-1}~Hz^{-1}}$ as
\begin{equation}
I_{\nu,\rm{jitter}}=\frac{8e^4}{3m_e^2c^3}\nu^{-(\zeta_p-1)},
\end{equation}
where $\zeta_p$ is the index determined by the turbulent energy
cascade and $c$ is the light speed. In this simplified case, we note
that the jitter radiation of a single electron is not related to the
electron Lorentz factor $\gamma$. The radiative flux of a single
electron can be simply estimated by
$I_{\nu,\rm{jitter}}ct_{\rm{cool}}$, where $t_{\rm{cool}}$ is the
radiative cooling timescale of relativistic electrons.

The process of inverse Compton scattering can be calculated by a
standard recipe \citep{rybicki79}. The SSC radiation has been fully
discussed as well \citep{chiang99}. In principle, our JSC process
can follow the same SSC calculation procedure. The emission flux
density in the unit of $\rm{erg~s^{-1}~cm^{-3}~Hz^{-1}}$ is
\begin{equation}
j_{\nu,\rm{jsc}}=8\pi
r_0^2ch\int_{\nu_{0,\rm{min}}}^{\nu_{0,\rm{max}}}\int_{\gamma_{\rm{min}}}^{\gamma_{\rm{max}}}
N(\gamma)f(\nu/4\gamma^2\nu_0)n_{\rm{ph}}(\nu_0)d\nu_0d\gamma,
\end{equation}
where $f(x)=x+2x^2\rm{ln}x+x^2-2x^3$ for $0<x<1$, $f(x)=0$ for
$x>1$, and $x\equiv \nu/4\gamma^2\nu_0$. The Thomson scattering
section is $\sigma_T=8\pi r_0^2/3=6.65\times 10^{-25}~\rm{cm^2}$.
$n_{\rm{ph}}(\nu_0)$ is the number density of seed photons, and it
can be easily calculated from the jitter radiation as
$n_{\rm{ph}}(\nu_0)=t_{\rm{cool}}\int (I_{\nu,\rm{jitter}}/h\nu)
d\nu \int N(\gamma)d\gamma$, $N(\gamma)$ is the electron energy
distribution. $\nu_{0,\rm{min}}$ and $\nu_{0,\rm{max}}$ are the
lower and upper limits of jitter radiative frequency, respectively.
$\gamma_{\rm{min}}$ and $\gamma_{\rm{max}}$ are the lower and upper
limits of relativistic electron Lorentz factor, respectively.

In general, the electron energy distribution can be given as a power
law: $N(\gamma)\propto \gamma^{-p}$, where $p=2.2$. In this paper,
the turbulent process is one of the vital points for GRB prompt
emission and particle acceleration. As mentioned in Section 1, due
to stochastic acceleration, the Maxwellian function of electron
energy distribution can be obtained. Here, we follow the description
of electron energy distribution given by \citet{giannios09} as
\begin{equation}
N(\gamma)=C\gamma^2\rm{exp}(-\gamma/\Theta)/2\Theta^3
\end{equation}
for $\gamma\le \gamma_{\rm{nth}}$ and
\begin{equation}
N(\gamma)=C[\gamma^2\rm{exp}(-\gamma/\Theta)/2\Theta^3](\gamma/\gamma_{\rm{nth}})^{-p}
\end{equation}
for $\gamma > \gamma_{\rm{nth}}$, where $C$ is the normalization
constant, $\gamma_{\rm{nth}}$ is the connection number between the
Maxwellian and power law components, and $\Theta= kT/m_ec^2$ is a
characteristic temperature. We use this mixed thermal-nonthermal
electron energy distribution to calculate jitter/JSC radiation. In
the case of $\gamma_{\rm{min}}=\gamma_{\rm{nth}}$, the mixed
thermal-nonthermal distribution is reduced to a pure power-law
distribution.

\subsection{Jet-in-jet Scenario}
We draw a sketch to illustrate the ``jet-in-jet" scenario, as shown
in Figure 1. The term of ``jet-in-jet" means that those
microemitters radiating as minijets are within the bulk jet.
\citet{giannios10} proposed an ``off-axis" parameter $\alpha$
defined by $\theta_j=\alpha/\Gamma_j$, where $\Gamma_j$ is the bulk
Lorentz factor of the jet launched by GRB and $\theta_j$ is the
related view angle. The gross Lorentz factor can be derived as
$\Gamma=2\Gamma_j\Gamma_e/\alpha^2$, and $\Gamma_e$ is the Lorentz
factor of the minijet. In our work, because these minijets point
randomly in the bulk jet but all of them move with a general
turbulent velocity, we use $\Gamma_e\sim \Gamma_t$, and
$\Gamma_t=10$ is the turbulent Lorentz factor adopted by
\citet{narayan09}.

The possibility of observing these minijets can be estimated by
$P=2\pi\int_0^\theta \rm{sin}\theta'
d\theta'/4\pi=\theta^2/4=1/4\Gamma^2$. The observed flux is $\nu
f(\nu)=P\delta^{2+w}\nu'f(\nu')$, where $\nu'f(\nu')$ is the flux
calculated in the GRB shell frame, $w$ is the spectral index, and
$\delta$ is the Doppler factor. Here, we take $w=1$ and $\delta\sim
\Gamma$.

In the GRB shell frame, the microemitter has the length scale of
$l_s=\gamma ct_{\rm{cool}}$. The total number of microemitters
within the fireball shell is $n=4\pi R^2\delta_s/l_s^3$, where
$R\sim 10^{13}$ cm is the fireball radius and
$\delta_s=ct_{\rm{cool}}$ is the thick of the shell. The length
scale of the turbulent eddy is $l_{\rm{eddy}}\sim R/\Gamma$
\citep{narayan09}. We can define a dimensionless scale as
$n_l=l_{\rm{eddy}}/l_s$. Therefore, we sum up the contributions of
the microemitters within the turbulent eddy and obtain the total
observed duration of GRB emission as $T=n_lnP\Gamma t_{\rm{cool}}$.
We calculate these parameters in the next subsection.

\subsection{High-Energy Emission of GRB 100728A}
To quantitatively study the jitter/JSC process, we use GRB 100728A
as an example. GRB 100728A was detected by {\it Swift}/XRT (0.3-10
keV), {\it Swift}/Burst Alert Telescope (BAT, 15-150 keV), {\it
Fermi}/Gamma-ray Burst Monitor (10-1000 keV), and {\it Fermi}/LAT
(above 100 MeV). The spectrum observed by theb BAT and XRT can be
well fitted by the Band function with spectral index $\alpha=-1.06$
and $\beta=-2.24$ and peak energy $E_{\rm{pk}}=1.0$ keV
\citep{abdo11}. This spectral function can be extrapolated to the
GeV band. The X-ray emission was dominated by a series of bright
X-ray flares with a maximum rate above 800 $counts~s^{-1}$. In the
time interval of strong X-ray flare activity, a significant GeV
emission was detected by {\it Fermi}/LAT. We take these
observational data from \citet{abdo11} and plot them in Figure 2.

We perform the jitter/JSC calculations to reproduce the
multiwavelength spectrum of GRB 100728A and also plot the results in
Figure 2. We adopt $\Gamma_j=100$, $\Gamma_t=10$, and ``off-axis"
parameter $\alpha=1$. $C=1.7\times 10^{10} ~\rm{cm^{-3}}$ is the
value of the electron number density in the relativistic shock.
Using the spectral index determined by the energy spectrum of
turbulent flow, we can reproduce X-ray and prompt emissions of GRB
100728A through the jitter mechanism. Here, $\zeta_p=2.24$ is in the
theoretical range of the turbulent energy cascade \citep{she94}.

With the electron energy distribution presented by Equations (3) and
(4), fixing $\gamma_{\rm{min}}=100$, $\gamma_{\rm{max}}=10^6$, and
$\gamma_{\rm{nth}}=10^3$, we can obtain the JSC result. We adopt
$\Theta=200$, which corresponds to the plasma temperature above
$10^{12}$ K. The radiative cooling timescale $t_{\rm{cool}}$ is
given as $2.2\times 10^{-8}$ s (see the calculation below). The
frequency $\nu_{0,\rm{max}}=4.2$ keV is also considered. It
indicates that X-ray emission, which is dominated by X-ray flares in
this case, provides enough target photons for the relativistic
electron scattering. For comparison, using a pure power-law electron
energy distribution ($\gamma_{\rm{min}}=\gamma_{\rm{nth}}=10^3$), we
also calculate the JSC result which is shown in Figure 2. From all
of the results mentioned above, we confirm that the JSC mechanism
with a mixed thermal-nonthermal electron energy distribution is one
possible origin of GRB 100728A GeV emission detected by {\it
Fermi}/LAT.

The jitter/JSC calculations in this work for reproducing the
multiwavelength spectrum of GRB 100728A are strongly dependent on
some of the parameters mentioned above. For example, the ``off-axis"
parameter $\alpha$ includes a wide range as $0<\alpha<\Gamma_j$. On
the other hand, as shown in Figure 2, the two observational data
points have large error bars. Despite these uncertainties, some
additional physical components can modify the hydrodynamics and
radiative spectrum of GRB. For instance, we may further consider the
gamma-ray photon annihilation as $\gamma\gamma$\ding{213}$e^+e^-$
and the $e^+e^-$ cooling as $e^+e^-$\ding{213}$\gamma\gamma$.
$\gamma\gamma$ opacity implies a minimum bulk Lorentz factor
\citep{nakar07}. \citet{hascoet11} gave a detailed study on the
consequences of gamma-ray photon annihilation. Although these topics
are out of the scope of our paper, we note the possibility that the
jitter/JSC production has minor differences with the
multi-wavelength observation.

Because the random and small-scale magnetic field can be generated
by the turbulent energy spectrum ($B^2=\int F(k)dk$, $F(k)\propto
k^{-\zeta_p}$; see Mao \& Wang 2011), in this work, we obtain
$B=1.0\times 10^6$ G. The cooling timescale of relativistic
electrons for the radiation of jitter and JSC is
$t_{\rm{cool}}=3m_ec/4\sigma_T\gamma(U_B+U_{\rm{ph}})$, where $U_B$
is the energy density of magnetic field and $U_{\rm{ph}}$ is the
energy density of radiation field. In the case of GRB 100728A, as
$U_B\gg U_{\rm{ph}}$, the cooling timescale is dominated by the
jitter radiation:
\begin{equation}
t_{\rm{cool}}=\frac{6\pi m_ec}{\sigma_T\gamma B^2}=2.2\times
10^{-8}(\frac{\gamma}{3.6\times 10^4})^{-1}(\frac{B}{1.0\times 10^6
~\rm{G}})^{-2}~\rm{s},
\end{equation}
where $\gamma=3.6\times 10^4$ is the average value of the electron
Lorentz factor obtained from \citet{giannios09}.

We use the reference value $t_{\rm{cool}}=2.2\times 10^{-8}$ s as
the radiative cooling timescale, which is much shorter than the
observed GRB duration. We expect that those fast-variability pulses
shown in the GRB prompt emission/X-ray flare are produced by the
extremely short-time activities of the microemitters. The gross
profile \citep{norris05} with a long-time duration of prompt
emission/X-ray flare is the superimposing of those fast-variability
pulses. We can further quantify the parameters mentioned in Section
2.2. In the shell frame, the length scale of the microemitters is
\begin{equation}
\l_s=\gamma ct_{\rm{cool}}=2.2\times10^6(\frac{B}{1.0\times
10^6~\rm{G}})^{-2}~\rm{cm}.
\end{equation}
The total number of microemitters within the shock shell of the
thick $\delta_s=ct_{\rm{cool}}$ is
\begin{equation}
n=\frac{4\pi R^2\delta_s}{l_s^3}=7.8\times
10^{10}(\frac{R}{10^{13}~\rm{cm}})^2(\frac{\gamma}{3.6\times
10^4})^{-2}(\frac{B}{1.0\times 10^6~\rm{G}})^4.
\end{equation}
The length scale of the turbulent eddy can be estimated as
\begin{equation}
\l_{\rm{eddy}}=\frac{R}{\Gamma}=5.0\times
10^{9}(\frac{\alpha}{1.0})^2(\frac{R}{1.0\times
10^{13}~\rm{cm}})(\frac{\Gamma_j}{100})^{-1}
(\frac{\Gamma_t}{10})^{-1}~\rm{cm}.
\end{equation}
With the definition $n_l=l_{\rm{eddy}}/l_s$, we sum up the
contributions from all of the microemitters in the turbulent eddy
and obtain the duration of the GRB prompt emission:
\begin{equation}
T=n_lnP\Gamma
t_{rm{cool}}=460(\frac{\alpha}{1.0})^4(\frac{R}{1.0\times
10^{13}~\rm{cm}})^3(\frac{\gamma}{3.6\times
10^4})^{-3}(\frac{B}{1.0\times
10^6~\rm{G}})^4(\frac{\Gamma_j}{100})^{-2}
(\frac{\Gamma_t}{10})^{-2}~\rm{s}.
\end{equation}
This calculated timescale is roughly consistent with the observed
GRB duration.

Finally, as \citet{honda05} and \citet{honda09} studied particle
acceleration in the random and small-scale magnetic field, we adopt
their result to calculate the acceleration timescale of relativistic
electrons for the GRB prompt emission:
\begin{equation}
t_{\rm{acc}}=4.0\times
10^{-12}(\frac{E}{\rm{MeV}})^2(\frac{B}{1.0\times
10^6~\rm{G}})^{-2}(\frac{l_{\rm{eddy}}}{5.0\times 10^9~
\rm{cm}})^{-1}(\frac{U}{0.1c})^{-2}~\rm{s},
\end{equation}
where upstream speed is $U\sim 0.1c$. After comparing the cooling
and acceleration timescales, and assuming $\gamma=3.6\times 10^4$
and $l_{\rm{eddy}}=5.0\times 10^9~\rm{cm}$, we obtain
$t_{\rm{acc}}\le t_{\rm{cool}}$ below 100 MeV. This indicates that
the particle acceleration is effective for the jitter mechanism.

\section{Conclusions and Discussion}
The gamma-ray emission of GRB 080319B provided evidence of the
relativistic turbulent process \citep{kumar09}. It was already
pointed out by \citet{nakar02} that the GRB temporal profile
contains many fast-variability pulses. These observational issues
give us general hints into the consideration of jitter radiation in
a turbulence-generated magnetic field. Using particle-in-cell (PIC)
simulations, \citet{nishikawa09} found that a partially developed
hydrodynamic-like shock structure can be created when the jet
propagates into an unmagnetized plasma. The synthetic radiation
spectra were also extracted from PIC simulations \citep{sironi09,
fre10}. Moreover, \citet{mizuno11} performed relativistic
magnetohydrodynamic simulations of a relativistic shock propagating
through an inhomogeneous medium. It was shown that the postshock
region becomes turbulent and the magnetic field is strongly
amplified. In our work, the magnetic field generated by relativistic
turbulence has a sub-Larmor length scale. We further suggest that
the ``jet-in-jet" scenario is causing the relativistically
counterstreaming plasma mentioned above.

In particular, \citet{reynolds10} and \citet{reynolds11}
comprehensively studied jitter radiation spectra that are dependent
on the properties of anisotropic magnetic turbulence. The radiation
spectra are also strongly affected by the spatial distribution of
the magnetic field. In our work, the magnetic field and radiation
are in the framework of the one-dimensional case. This prevents us
from further investigating the topologies of the turbulent and
magnetic fields in detail. However, we can compare our result and
the result of \citet{reynolds10} and \citet{med11} in the
one-dimensional case. For example, in the study of \citet{med11},
the isotropic turbulence has the form of $f(k)\sim k^{-2\beta}$
after the nonlinear evolution and the magnetic field is $B^2=f(k)$,
while it is $B^2=k^{\zeta_p-1}$ in our work. Therefore, in the
one-dimensional case, we obtain $\beta=(\zeta_p-1)/2$. Moreover,
from the study of \citet{med11}, we know that $f(k)$ is strongly
related to the topology of turbulence and the view angle $\theta$.
The jitter parameter is given as $K=eBl_{\rm{cor}}/mc^2$, where
$l_{\rm{cor}}$ is the correlation scale, so we can clearly see that
the jitter parameter is also strongly related to the topologies of
the turbulent and magnetic fields.

Although the GRB emission from about 10 keV up to 10 GeV can usually
be fitted by a single radiation process (see Abdo et al. 2009 for
the case of GRB 080916C and Ackermann et al. 2010a for the case of
GRB 090217A), sometimes an additional component is required to
completely explain the GRB GeV emission. At present, GRB 090510 and
GRB 100728A are two sources that have been studied by using
published data sets from the simultaneous observations of {\it
Swift} and {\it Fermi}. Besides GRB 100728A, GRB 090510 is another
interesting source used to examine the jitter and JSC processes.
{\it Fermi}/LAT detected the emission above 100 MeV
\citep{ackermann10}. \citet{depasquale10} built the multiwavelength
spectral energy distribution (SED) from simultaneous observations of
{\it Swift} and {\it Fermi}. Synchrotron and SSC were proposed to be
the origins of the keV-MeV emission and an additional component
above the GeV band, respectively \citep{ackermann10b}, and a large
bulk Lorentz factor $\Gamma>500-1000$ was required. However, it is
difficult to apply the simple model presented in this paper for this
burst because of two reasons. (1) A hard power-law component
dominates the emission both below 20 keV and above 100 MeV
\citep{ackermann10b}. The JSC process can be adopted to explain the
emission above 100 MeV, but it cannot explain the emission below 20
keV. (2) There are no BAT data shown in the multiwavelength SED
\citep{depasquale10}. The spectral data in the energy range of
$0.3-10$ keV cannot well constrain the jitter slope if jitter
radiation is dominated from 1 keV to 100 MeV. Moreover, it is also
difficult to use the {\it Fermi}/LAT spectral data above 100 MeV
plotted with a wide-ranging confidence interval (see the butterfly
in Figure 3 of De Pasquale et al. 2010) to constrain the parameters
of the JSC model.

In this paper, we use the JSC mechanism to interpret the powerful
emission of GRB 100728A above the GeV band. In particular, we see
two extraordinary observational behaviors of GRB 100728A. (1) A
series of powerful X-ray flares (maximum rate is larger than 500
$counts~s^{-1}$) that occurred about 800 s after the burst trigger.
(2) The significant GeV band emission was detected by {\it
Fermi}/LAT in the same time interval as the X-ray flares. From this
observational evidence, we were able to examine the jitter and JSC
mechanisms. Using the jitter mechanism, we successfully reproduced
prompt and X-ray emissions of GRB 100728A. To use the JSC mechanism
to explain the high-energy emission of GRB 100728A above the GeV
band, the jitter photons in the X-ray band should be scattered by
the mixed thermal-nonthermal electrons to the GeV band. Meanwhile,
the ``jet-in-jet" scenario is also considered. Therefore, our model,
which includes all of the main points discussed above, such as
turbulence, the random and small-scale magnetic field, radiation,
and geometric structure of emission, is self-consistent.

We confirm through our calculation that the jitter and JSC processes
in the ``jet-in-jet" scenario are valid for the multiwavelength
emission of GRB 100728A without any further assumptions. However, as
shown in Figure 2, the JSC result has a minor deviation to the
observational data above 100 MeV. The effect of the $\gamma\gamma$
opacity on the output spectrum is mentioned in Section 2.3. Here
some other physical components are generally proposed. Although the
bulk Lorentz factor of GRB jet $\Gamma_j=100$ and the Lorentz factor
of the minijet $\Gamma_e=10$ are given, the shock and emission
regions may have a complex structure, that can modify the radiation
spectrum. The radiative efficiency is also an important issue. The
relation between the cooling time of the internal shock and the
shock Lorenz factor was investigated by \citet{med09}. Moreover,
instead of a constant density of shock in our case, a certain
structured density profile may be involved. We speculate that the
final output radiation is the superposition of multicomponent
contributions.

From Figure 2, we see that the JSC radiation above the GeV band is
related to the electron energy distribution. The powerful emission
above the GeV band can be obtained if the seed photons are scattered
by the electrons with a mixed thermal-nonthermal distribution. The
weak GeV emission may have a flux value below the threshold of what
an observing instrument can detect, if the seed photons are
scattered by electrons with a purely nonthermal power-law
distribution. In our former research \citep{mao11}, the acceleration
timescale is larger than the cooling timescale above 100 MeV. Thus,
the jitter radiation does not work and we do not expect more
GeV-GRBs. In this paper, we further explore GRB detection above the
GeV band by {\it Fermi}/LAT that is also strongly dependent on
particle acceleration. Some other interesting suggestions, such as
the upscattered cocoon model \citep{toma09} and the external inverse
Compton model \citep{he11}, may also be important in explaining GRB
detection by {\it Fermi}/LAT. We hope that more multiwavelength data
sets can be accumulated so that more penetrating studies can be
performed in the future\footnote{The simultaneous observations of
{\it Swift} and {\it Feimi}/LAT were performed on GRB 110625A
\citep{page11,gruber11,tam11} and GRB 110731A
\citep{oates11,bregeon11,gruber11}. We hope that future published
data can provide more constraints on our model.}.



\acknowledgments

We thank the referee for the instructive suggestions. This work is
supported by the KASI Fellowship program, the National Natural
Science Foundation of China 11173054, the National Basic Research
Program of China (973 Program, 2009CB824800), and the Policy
Research Program of the Chinese Academy of Sciences (KJCX2-YW-T24).

\clearpage



\begin{figure}
\includegraphics[angle=0,scale=2.65]{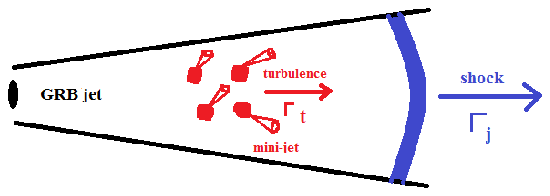} \caption{Illustration of the ``jet-in-jet"
scenario. The turbulent process occurs after shock propagation. The
bulk Lorentz factor of the GRB jet is $\Gamma_j$. The microemitter
with its minijet has the Lorentz factor $\Gamma_e$, and
$\Gamma_e\sim \Gamma_t$, where $\Gamma_t$ is the Lorentz factor of
turbulence. \label{fig1}}
\end{figure}

\clearpage

\begin{figure}
\includegraphics[angle=270,scale=0.65]{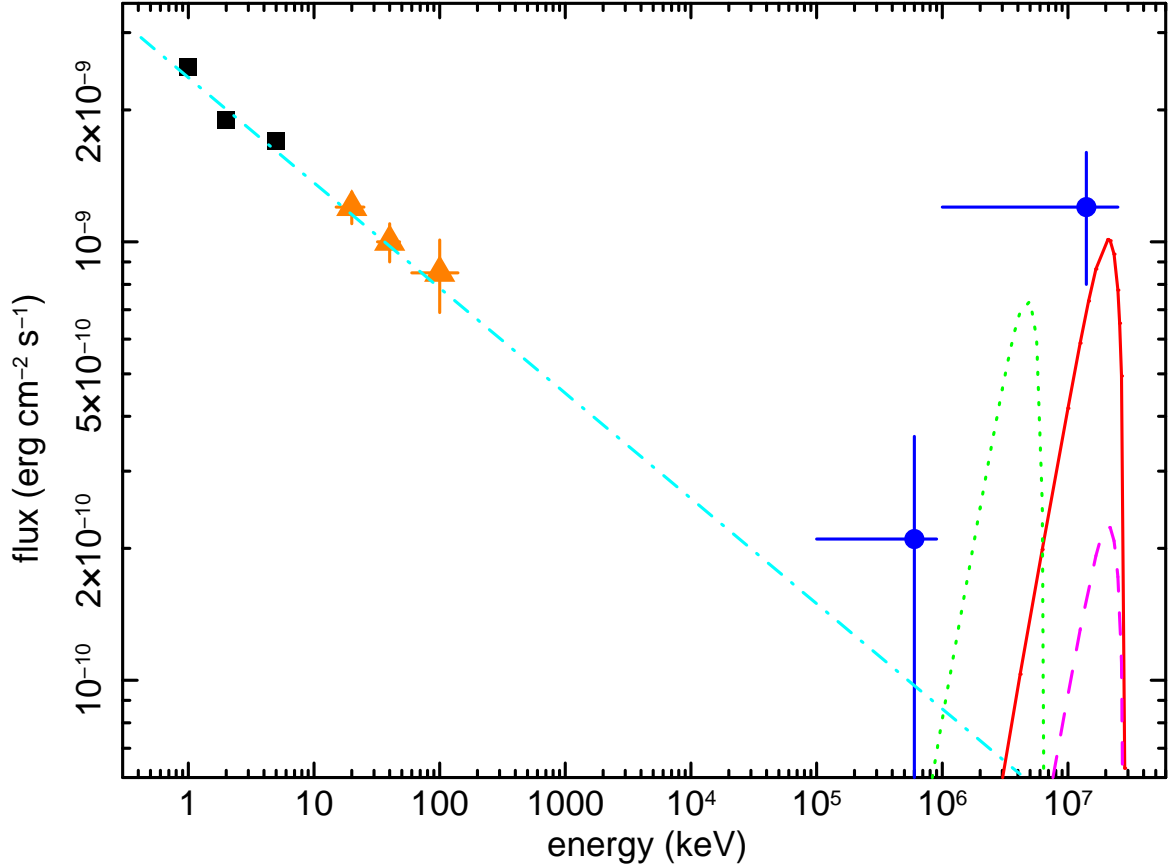}
\caption{Jitter and JSC radiation of GRB 100728A. The observational
data from \citet{abdo11} are denoted by solid squares (XRT data),
solid triangles (BAT data), and solid circles (LAT data). The result
of jitter radiation with the spectral slope 0.24 is denoted by a
dash-dotted line. The JSC results calculated using a mixed
thermal-nonthermal electron energy distribution (given by Equations
(3) \& (4)) are shown: a solid line denotes the JSC result in the
case of $\gamma_{\rm{max}}=10^6$ and a dotted line denotes the JSC
result in the case of $\gamma_{\rm{max}}=5\times 10^5$. For
comparison, the JSC result calculated using a power-law electron
energy distribution ($\gamma_{\rm{min}}=\gamma_{\rm{nth}}=10^3,
\gamma_{\rm{max}}=10^6$) is denoted by a dashed line. \label{fig2}}
\end{figure}

\clearpage

\end{document}